\documentstyle[aps,prb]{revtex}
\begin{document}
\title{Acoustic attenuation in a type-II superconductor at high magnetic fields }
\author{G. M. Bruun, V. Nikos Nicopoulos, N. F. Johnson}
\address{Department of Physics,
Clarendon Laboratory,
University of Oxford,
Oxford OX1 3PU,
England}
\date{\today}
\maketitle

\begin{abstract}
We have calculated the longitudinal acoustic attenuation in a type II superconductor in high magnetic 
fields within a mean-field BCS theory. We predict two new features in the corresponding 
attenuation signal as compared to that of the Meissner state. Our analytical calculations 
predict the existence of oscillations in the attenuation as the external magnetic field
 is varied- this effect is associated with the  Landau level structure of the electron states
and is analogous to  the 
well-known de Haas van Alphen oscillations in the mixed state. The attenuation 
directly probes the quasiparticle energies; the presence of gapless points in the
 quasiparticle spectrum, which is  characteristic of type-II superconductors at high magnetic
 fields, shows 
up in  the frequency $\omega$ and temperature $T$ dependence of the attenuation in 
the limit of low $\omega$ and and low $T$ respectively. At low $T$
 there is no analogue to the discontinuity in the attenuation observed in the 
Meissner state when $\hbar\omega=2\Delta$, where $\Delta$ is the quasiparticle energy
 gap.  This result opens up the possiblity of experimentally determining the existence 
and nature of the gapless points in the quasiparticle spectrum of a type-II superconductor 
in high magnetic fields.

\end{abstract}
\


\section{Introduction}
The problem of understanding type-II superconductivity in  high magnetic fields has been a
 lively topic for some years now. When $k_BT < \hbar\omega_c$,  where 
 $\omega_c=eH/mc$ is the cyclotron frequency, the Landau level 
quantization of the electron levels become important. This gives rise to a number of 
interesting effects.~\cite{Rasolt} One such effect is the existence of magnetic oscillations 
in the free energy of the superconductor in the mixed state. These de Haas-van Alphen 
(dHvA) oscillations was observed over 20 years ago~\cite{Graebner} and have since been
 observed in a number of materials. One important theoretical consequence of the 
Landau level quantization is the presence of gapless points in the quasiparticle spectrum.  
Theoretical investigations have shown~\cite{Big Mac,Zlatko} that the quasiparticle spectrum 
of the mixed state in high magnetic fields is characterized by a set of gapless points in the 
magnetic Brillouin zone (MBZ). These gapless point leads to an algebraic temperature 
dependence of the thermodynamic functions and an algebraic voltage dependence 
in the tunneling conductance.~\cite{Zlatko} 
It has been suggested~\cite{Dukan} that the existence of the 
dHvA oscillations in the mixed state is a  consequence of these gapless points.  It turns out
 however, that it is rather difficult  to develop a simple, yet consistent theory for the 
dHvA-oscillations since  both the oscillatory behaviour of the ground state 
energy and the gapless nature of the quasiparticle energies need to be taken into 
account.~\cite{Bruun} 
The interpretation of the experimental results for the dHvA-oscillations is consequently 
somewhat unclear and does not, in our opinion, give a completely unambiguous signature 
for the presence of gapless modes.

In this paper we consider the attenuation of longitudinal acoustic waves in the mixed state 
where the order parameter is assumed to form a vortex lattice. Since the absorption of the 
phonons is due to quasiparticle excitations, the experiment directly probes the 
quasiparticle density of states. For clean materials, theoretical results have proved rather
 difficult to obtain since an expansion in powers of the order parameter does not 
converge~\cite{Maki} and  
the semiclassical approximation leads to unphysical results.~\cite{Scharnberg} Since we
 consider 
the superconductor at high magnetic fields, it is crucial to take into account the Landau level 
quantization of the electronic levels. It turns out that by including this effect, one 
avoids the difficulties encountered in the semiclassical approximation as predicted by 
Scharnberg.~\cite{Scharnberg} We calculate analytically the 
attenuation in two limits: $k_B T \ll\hbar \omega$ and $k_B T \gg\hbar \omega$ 
where $\omega$ is the frequency of the sound wave and $T$ is the temperature. 
We predict that the attenuation will be an oscillatory function of the external magnetic 
field due to the Landau level structure, in analogy with the usual dHvA oscillations. 
The frequency and temperature dependence of the attenuation is  determined 
by the existence and the nature of the gapless points in the quasiparticle spectrum for 
 $k_B T \ll\hbar \omega$ and $k_B T \gg\hbar \omega$ respectively. This in principle 
gives an experimental tool for probing the nature of the quasiparticle energies in the mixed
state. Our analytical theory is supported by  essentially exact numerical calculations for the 
attenuation. 

As the analysis in this article will show, the existence of the gapless points in the quasiparticle 
spectrum has significant consequences on the longitudinal sound attenuation. We would 
therefore expect that this should also hold for systems with intrinsic gapless points even for 
no external field such as high-temperature superconductors (high-$T_c$'s), which seems
 to have a  $d$-wave gap symmetry. Hence a similar  analysis of the acoustic attenuation in these 
high-$T_c$'s as the one presented in this paper for type-II superconductors in high magnetic fields 
would be very interesting, and we will return to this in a future publication.

\section{General Theory}
We consider a weak-coupling superconductor in three dimensions (3D) described within mean 
field theory by the following Hamiltonian:
\begin{equation} \label{hamilton}\hat{H}=\sum_{\sigma}\int d{\mathbf{r}}\,
 \psi_{\sigma}^{\dagger}({\mathbf{r}})
\left( \frac{({\bf p}-\frac{e}{c}{\bf A})^2}{2m} -\mu\right)\psi _{\sigma}({\mathbf{r}}) + \int d{\mathbf{r}}\,[\Delta({\mathbf{r}})\psi_{\uparrow}^{\dagger}({\mathbf{r}})
\psi _{\downarrow}^{\dagger}({\mathbf{r}})+c.c]
 \end{equation}
where the order parameter is defined as 
 $\Delta({\mathbf{r}})=g\langle \psi_{\uparrow}({\mathbf{r}})\psi_{\downarrow}({\mathbf{r}})\rangle$
, $g$ is the coupling strength and $\mu$ is the chemical potential. 
Our theory describes  longitudinal ultrasonic waves in clean samples for 
high magnetic fields.   We have for simplicity confined our theory to the case when 
 ${\mathbf{q}} \parallel {\mathbf{H}}$ where ${\mathbf{q}}$ is the wavevector of the sound 
wave and ${\mathbf{H}}$ is the magnetic field.  This simplifies our calculations since the 
coupling between the collective modes associated with fluctuations of the order parameter and 
the longitudinal phonons can be neglected to a good approximation.~\cite{Caroli,Dominguez}
The attenuation of sound waves is given by the imaginary part of the retarded phonon 
self energy. To lowest order in the electron-phonon coupling we 
get for the attenuation $\alpha(\omega)$:~\cite{Mahan}
\begin{equation} \label{start}\alpha(\omega) \propto -\omega Im\left\{ D^{R}({\mathbf{q}},\omega) 
\right\}\end{equation}
where $ D^{R}({\mathbf{q}},\omega)$ is the retarded density-density correlation function
 \begin{equation}  iD^{R}({\mathbf{x}},{\mathbf{x}}',t-t')=\langle[\tilde{n}({\mathbf{x}},t),
\tilde{n}({\mathbf{x}}',t')]\rangle\theta(t-t')\end{equation}
 and $\tilde{n}({\mathbf{x}},t)=\hat{n}({\mathbf{x}},t)-\langle\hat{n}({\mathbf{x}},t)\rangle$ 
is the operator describing density fluctuations. 
This function is evaluated as the analytical continuation of 
the thermal correlation function. We will in this paper for simplicity ignore the electronic 
Zeeman effect. The effect  of a finite spin splitting on the magnetic oscillations is well 
understood both in the normal state~\cite{Schoenberg} and in the mixed state.~\cite{Bruun} 
Hence ignoring the spin splitting we obtain
\begin{eqnarray} D^{R}({\mathbf{q}},\omega)=\frac{2k_{B}T}{V_{cell}}\sum_{\omega_
{\nu}}
\int_{cell} d^2{\mathbf{r}} \int d^3{\mathbf{r}}' e^{{\mathbf{qr}}'}\left[ 
G({\mathbf{r}},{\mathbf{r}}',\omega_{\nu})G({\mathbf{r}}',{\mathbf{r}},\omega_{\nu}-
\omega_{\gamma}) \right.\nonumber \\ \left. \left.
-F^{\dagger}({\mathbf{r}},{\mathbf{r}}',\omega_{\nu})F({\mathbf{r}}',{\mathbf{r}},
\omega_{\nu}-\omega_{\gamma}) \right]\right|_{i\omega_{\gamma}\rightarrow \omega 
+i\delta} \end{eqnarray}
where $G({\mathbf{r}},{\mathbf{r}}',\omega_{\nu})$ and 
 $F({\mathbf{r}}',{\mathbf{r}},\omega_{\nu})$ are the one-particle Green's functions for the 
superconductor. The symbol $\int_{cell} d^2{\mathbf{r}}$ implies  integration  over
 one vortex lattice cell in the $x\!-\! y$ plane whereas $\int d^3{\mathbf{r}}'$ means 
integration over the whole crystal. By 
treating the electron-phonon matrix element as an overall factor in our formalism 
(i.e. its frequency dependence $\omega$ is included as a prefactor in Eq.(\ref{start})) we 
have made use of the fact
 that the screening for longitudinal modes is essentially the same as in the normal 
phase.~\cite{Schrieffer} Since the order parameter is 
assumed to form a vortex lattice, we can use a set of single particle states 
 $\phi_{n {\mathbf{k}}k_z}({\mathbf{r}})$ characterized by a Landau level index $n$, a 
wavevector 
 ${\mathbf{k}}=(k_x,k_y)$ in the MBZ, and a wavevector $k_z$ along the $z$-direction. 
These basis states were introduced  
 by Norman \textit{et al}.~\cite{Big Mac} In this basis there is no mixing between 
different ${\mathbf{k}}$'s since the sound wave travels along the $z$-direction. Expanding the 
Green's functions in this basis and using the ${\mathbf{k}}$-space symmetry of the 
problem due to the presence of the vortex lattice,~\cite{Big Mac} we obtain 
 the following expression for the ultrasonic attenuation in the mixed state of a type-II 
superconductor:
\begin{eqnarray} \label{master}
\alpha({\mathbf{q}},\omega)&\propto& \frac{\omega}{L_xL_y}\sum_{n n'}
\sum_{{\mathbf{k}}}\int dk_z\left\{[f(E)-f(E')]  [\delta(E'-E-\omega) 
(|U'\wedge U|^2- \right. \nonumber \\
 && \left. -V'\wedge V^*\,U'^*\wedge U)-
 \delta(E-E'-\omega)(|V'^*\wedge V|^2-U'^*\wedge U\, V'\wedge V^*)]  \right.\nonumber \\
&&\left. +[1-f(E')-f(E)]\delta(E'+E-\omega)(|U'\wedge V|^2+V'\wedge U\, U'^*\wedge V^*)
 \right\}.\end{eqnarray}
Here $U\equiv U^{n}_{{\mathbf{k}}k_z}({\mathbf{r}})$,
 $V\equiv V^{n}_{{\mathbf{k}}k_z}({\mathbf{r}})$,
 $U'\equiv U^{n'}_{{\mathbf{k}}k_z+q}({\mathbf{r}})$ and 
 $V'\equiv V^{n'}_{{\mathbf{k}}k_z+q}({\mathbf{r}})$ are the Bogoliubov functions. The 
quasiparticle energies are given by $E\equiv E_{k_z}^{n}({\mathbf{k}})$ and 
 $E' \equiv E_{k_z+q}^{n'}({\mathbf{k}})$; $f(E)=1/(\exp(E/k_BT)+1)$ and the 
$\wedge$-product means integration 
over the $xy$-plane. The quasiparticle energies and associated wavefunctions can be found 
by solving the corresponding Bogoliubov-de Gennes (BdG) equations.~\cite{deGen} 
We have set up a program that solves the BdG equations self-consistently 
in 3D such that we get the Bogoliubov functions as a function of $H$. The method 
for solving these equations numerically in 2D is described in detail 
elsewhere.~\cite{Big Mac,Bruun}The extension to 3D is straightforward although it is 
computationally significantly more demanding.  
 We choose the normal state dispersion law along the $z$-direction to be 
either the plane wave form $\epsilon(k_z)=k_z^2/2m$, or more suitable for layered 
structures, the tight-binding form  $\epsilon(k_z)=t\cos(k_za_z)$ where $a_z$ is the 
distance between the planes.

In order to develop an analytical theory for the attenuation we need to make some 
approximations. Near the upper critical field $H_{c2}$ and for quasiparticle levels close to the 
Fermi energy we can, as a first approximation, ignore the off-diagonal pairing (i.e. the 
so-called diagonal approximation). The 
quasiparticle energies are then given by~\cite{Nicopoulos}
\begin{equation} \label{Eapprox}
E_{k_z}^n({\mathbf{k}})=\sqrt{\xi_n(k_z)^2+|\Delta({\mathbf{k}})|^2}
\end{equation} 
and the corresponding Bogoliubov functions are given by
\begin{equation} \label{UVapprox}
\left.\begin{array}{l}|U^{n}_{{\mathbf{k}}k_z}({\mathbf{r}})|^2\\
|V^{n}_{{\mathbf{k}}k_z}({\mathbf{r}})|^2\end{array}\right\}
=\frac{1}{2}(1\pm\frac{\xi_n(k_z)}{E_{k_z}^n({\mathbf{k}})})
|\phi_{n {\mathbf{k}}k_z}({\mathbf{r}})|^2.
\end{equation}
Here $\xi_n(k_z)=n\hbar\omega_c+\epsilon(k_z)-\mu$.
It should be noted that this approximation is only valid for energy levels close to the 
Fermi energy. Further away from the Fermi level there are degeneracies between 
electron and hole states belonging to different Landau levels, hence our approximation 
will eventually breake down. However, for low frequencies and temperatures one can show
 that only 
levels for which Eq.(\ref{Eapprox}) and Eq.(\ref{UVapprox}) hold will contribute to the damping 
described by Eq.(\ref{master}). A closer examination shows the requirement  for the diagonal 
approximation  to hold to be 
 $\max (k_BT,\omega)\,\raisebox{-0.4ex}{$\stackrel {<}{\sim}$}\,\hbar\omega_c/4$. 
 This follows because  mixing with hole levels become important for levels with 
$E\ge\hbar \omega_c/4$.~\cite{Bruun,Big Mac2}  Dukan 
\textit{et al.}~\cite{Zlatko,Nicopoulos} argued that the off-diagonal pairing does not change the
 qualitative behavior of the superconductor in a high magnetic field for fields not too far
 below $H_{c2}$, and that the quasiparticle spectrum remains essentially the same when the 
off-diagonal terms are included. From the diagonal approximation, it 
directly follows that there are gapless points in the MBZ (i.e $\Delta({\mathbf{k}})=0$).
 Even when the diagonal approximation breaks down, there will still be points in the MBZ 
where the gap vanishes. It seems the role of the off-diagonal terms is to shift the 
value of the Fermi momentum $k_{zf}$ where the gapless 
behaviour occurs away from its diagonal approximation value 
 $\epsilon(k_{zf})=0$.~\cite{Zlatko} Eventually, true gapped behaviour sets in when 
the superconducting order is strong enough to increase the energies of the quasibound 
states in the vortices above $\hbar \omega_c$.~\cite{Big Mac}  

 In Fig. 1 we have plotted the 
lowest quasiparticle energy, calculated numerically, along the $\Gamma-M$ direction
 in ${\mathbf{k}}$-space for two different values of 
the order parameter $\Delta({\mathbf{r}})$ at a low temperature. The two points 
 $\Gamma=(\pi/4a_x,\pi/2a_y)$ and $M=(3\pi/4a_x,\pi/2a_y)$ are two of the 
corners in an irreducible triangle reflecting the symmetry of the BdG-equations in 
 ${\mathbf{k}}$-space.~\cite{Big Mac} Here $a_x=l(\sqrt3\pi/2)^{1/2}$ as we have 
chosen a triangular symmetry for the vortex lattice, $a_y=l^2/a_x$  and 
 $l^2=\hbar c/eH$ is the magnetic
 length. We have chosen the size of the system such that $\pi/(2a_x\Delta k_x)=50$ where 
 $\Delta k_x=2\pi/Lx$ and $L_x$ is the extend of the system in the $x$-direction.
 The dashed lines give the
 diagonal approximation to the energies and the solid lines the exact numerical result. The 
two highest-lying curves are calculated with 
 $\langle \Delta ({\mathbf{k}}) \rangle\simeq 0.3\hbar \omega_c$ and the two lowest lying
 curves are with  $\langle \Delta ({\mathbf{k}}) \rangle\simeq 0.05\hbar \omega_c$. Here 
 $\langle \Delta ({\mathbf{k}}) \rangle$ means the ${\mathbf{k}}$-space average of the 
diagonal matrix element in the BdG equations. There are 
10 Landau levels within the pairing width and the dispersion law along the z-direction is 
 $\epsilon(k_z)=t\cos(k_za_z)$. The value of $k_z$ is chosen such that $\epsilon(k_z)=0$. 
 As can be seen, the diagonal approximation predicts two 
gapless points along this ${\mathbf{k}}$-space direction, both  with a linear dispersion law. 
For a small pairing parameter 
there is good agreement between the diagonal approximation and the full calculation. For
 a larger pairing parameter, i.e.  
 $\langle \Delta ({\mathbf{k}}) \rangle\simeq 0.3\hbar \omega_c$, the approximation is less 
precise- indeed the gapless points predicted by the diagonal approximation disappear in the 
full self-consistent calculation. This is due to the off-diagonal pairing which 
become increasingly 
important as the pairing interaction increases deeper into the mixed state. The gapless points 
are now at a different value of $k_z$. We expect 
our theory  to be valid reasonably close to the transition line such that we can ignore the 
off-diagonal pairing. As will be shown later, the major contribution actually comes from the
 gapless points where the diagonal approximation is most valid.
We will now calculate the attenuation in two limits using this approximation.

\section{Low frequency}

We will first treat  the case $\omega \ll k_BT$. In this limit we can focus on the first two delta 
functions in Eq.(\ref{master}) which describe the scattering of a quasiparticle. 
 We will outline the calculation for the dispersion law $\epsilon(k_z)=k_z^2/2m$. 
Making the approximation $\xi'/E'\simeq \xi/E$ which is valid for  $\hbar \omega\ll k_BT$
  the first two terms of Eq.(\ref{master}) become:
\begin{equation}
\alpha({\mathbf{q}},\omega)=-\frac{\omega^2 m}{\hbar L_xL_yq}\sum_{n}
\sum_{{\mathbf{k}}}\int dk_z\delta(k_z-k_z^*(n,{\mathbf{k}}))\partial_E f(E_{k_z}
^{n}({\mathbf{k}}))
\frac{|\xi_{n}(k_z)|}{{E_{k_z}^{n}({\mathbf{k}})}}
\end{equation}
Here $k_z^*(n,{\mathbf{k}})$ is the solution to the equation
 $E_{k_z+q}^{n}({\mathbf{k}})-E_{k_z}^{n}({\mathbf{k}})=\omega$. 
 From the Poisson formula we have the identity 
\begin{equation}\sum_{n=-\infty}^{\infty}\partial_E f(E_{k_z^*}
^{n}({\mathbf{k}}))
\frac{|\xi_{n}(k_z^*)|}{{E_{k_z^*}^{n}({\mathbf{k}})}} =
\sum_{j=-\infty}^{\infty}\int dx^{2j\pi ix}\partial_E f(E_{k_z^*}
^x({\mathbf{k}}))\frac{|\xi_x(k_z^*)|}{{E_{k_z^*}^x({\mathbf{k}})}}
 \end{equation}
where 
 $\xi_x(k_z^*)=x\hbar\omega_c+\epsilon(k_z^*)-\mu=(x-n_f)\hbar\omega_c+\epsilon(k_z^*)$. 
Here $n_f\equiv \mu/\hbar\omega_c-0.5$ is the Landau level index at the chemical potential.
We have extended 
 the Landau level sum to go from $-\infty$ as the low levels do not contribute anyway.
Making the variable 
substitution $ \xi_x(k_z^*)=z\hbar\omega_c$ we end up with
\begin{equation} \label{pois}
\alpha({\mathbf{q}},\omega)=-\frac{\omega^2 m}{\hbar L_xL_yq}\sum_{{\mathbf{k}}}
\sum_{j=-\infty}^{\infty}e^{2\pi i j (n_f-\epsilon(k_z^*)/\hbar\omega_c)}
\int_{-\infty}^{\infty} dz e^{2\pi i j z}\partial_E f(E_z({{\mathbf{k}}}))
\frac{|\xi_z|}{E_z({{\mathbf{k}}})}
\end{equation}
 where $E_z({{\mathbf{k}}})=(\xi_z^2+\Delta({\mathbf{k}}))^{1/2}$.
We have approximated $k_z^*$ as the corresponding normal state solution (i.e.\ it is 
independent of $n$). This approximation which is only necessary when we calculate the 
oscillations of the attenuation (i.e.\ the terms in Eq.(\ref{pois}) with $j\neq0$),  should be 
good close to the gapless points and is
 equivalent to  ignoring any phaseshift in the oscillations due to the superconducting order.
 The 0'th harmonic is given by the $j=0$ term. We get 
\begin{equation}\label{kint}
\alpha({\mathbf{q}},\omega)_0=
\frac{2\omega^2 m}{\hbar^2\omega_cL_xL_yq}\sum_{{\mathbf{k}}}\frac{1}
{e^{\Delta({\mathbf{k}})/k_BT}+1}
\end{equation}
Using an ansatz of a formal similarity between a pure type II superconductor in high 
magnetic fields and a current-carrying superconductor Maki~\cite{MakiII} arrived at 
the following expression for the ratio $\alpha_S/\alpha_N$ between longitudinal attenuation
 in the mixed state and in the normal state:
\begin{equation}
\frac{\alpha_S}{\alpha_N}\simeq 1-\frac{\Delta}{2k_BT}
\end{equation} 
where $\Delta^2\equiv\langle|\Delta({\mathbf{r}})|^2\rangle$ is the real space average of the 
gap. As Maki used a semiclassical approximation he did only calculate the 0'th harmonic. To compare
 with this result we expand Eq.(\ref{kint}) to first order in $\Delta({\mathbf{k}})$. 
We get:
\begin{equation}
\frac{\alpha_S}{\alpha_N}=1-\frac{\langle|\Delta({\mathbf{k}})|\rangle_{{\mathbf{k}}}}{2k_BT}
\end{equation}
where $\langle|\Delta({\mathbf{k}})|\rangle_{{\mathbf{k}}}$ is the ${\mathbf{k}}$-space average 
of $|\Delta({\mathbf{k}})|$. We see that  our theory for the 0'th harmonic, which is exact within the 
diagonal approximation produces a term linear in $\Delta$ for the ratio $\alpha_S/\alpha_N$ as 
 does Maki's conjecture. This linear term is somewhat surprising as  the Gor'kov expansion 
of the Green's functions would seem to imply that the first correction term is quadratic in $\Delta$. 
However, even in the zero field BCS-state one obtains~\cite{Schrieffer} 
  $\alpha_S/\alpha_N=2/(\exp(\Delta/k_BT)+1)$ which cannot be expanded in $\Delta^2$. 
So our theory confirms Maki's ansatz of a linear correction term for the damping. Thus the 
non-perturbative linear term in the quasiparticle energy Eq.(\ref{Eapprox}) coming essentially 
from the degeneracy of the Landau levels shows up in the attenuation result, making a 
calculation based on the 
 Gor'kov expansion questionable. This is in contrast to the case of the expansion of the 
thermodynamic potential where contrary to earlier claims,~\cite{Bahcall} it can be proven that 
these non-perturbative terms cancel, thus making the Gor'kov expansion of the thermodynamic 
potential correct.~\cite{Bruun2}
Our result  substitutes Maki's real space average 
 $\Delta=(\langle|\Delta({\mathbf{r}})|^2\rangle)^{1/2}$ with the  ${\mathbf{k}}$-space average 
of $|\Delta({\mathbf{k}})|$. 

For conventional superconductors in the Meissner state we know that the 
finite gap for all ${\mathbf{k}}$ supresses the attenuation by a factor 
$2/(\exp(\Delta/k_BT+1)$. 
 We expect that the existence of gapless points will now change this result significantly. We
 therefore split the MBZ into two qualitatively different regions: the ``gapped'' region where
 we assume  $\Delta({\mathbf{k}})= \langle\Delta({\mathbf{k}})\rangle$ where 
 $\langle\Delta({\mathbf{k}})\rangle$ is 
the ${\mathbf{k}}$-space average of $\Delta({\mathbf{k}})$,  a the ``gapless'' region where 
we assume $\Delta({\mathbf{k}})= \gamma k^{\eta}$. We furthermore assume that this
latter  dispersion 
law holds for all $\Delta({\mathbf{k}}) \,\raisebox{-0.4ex}{$\stackrel {<}{\sim}$}\,k_BT$.
 This is a slight generalization 
of the model used by Dukan and  Te\v{s}anovi\'{c}~\cite{Dukan} in their theory for the 
dHvA oscillations. We assume that the gapped region takes up a fraction $\mathcal{F}$ of the 
MBZ. The contribution $\alpha({\mathbf{q}},\omega)_{0,gap}$ to the attenuation from the 
gapped region is then
\begin{equation}\label{gapped}
\alpha({\mathbf{q}},\omega)_{0,gap}=\frac{\omega^2 m^2}{\hbar^2qh}{\mathcal{F}}
\frac{2}{e^{\langle\Delta({\mathbf{k}})\rangle/k_BT}+1}
\end{equation}
where $h=\hbar 2\pi$.
As expected the attenuation from the gapped part of the spectrum is strongly supressed 
due to the $2/(\exp(\langle\Delta({\mathbf{k}})\rangle/k_BT)+1)$ factor. 
This result for the 0'th harmonic of the attenuation from the gapped part of the spectrum 
in the mixed state is the same as for the total attenuation in the Meissner state of a
 conventional superconductor. The qualitative new feature comes from the presence of the 
gapless points. Solving the 2-dimensional ${\mathbf{k}}$-space integral in Eq.(\ref{kint}) 
using the assumed dispersion law, we obtain  the contribution 
$\alpha({\mathbf{q}},\omega)_{gl}$ from the gapless part as
\begin{equation} \label{gapless}
\alpha({\mathbf{q}},\omega)_{0,gl}=Q_{gl}\frac{m\omega^2}{\hbar^2\omega_c q\pi}\left(\frac{k_BT}
{\gamma}\right)^{2/\eta}\frac{1-2^{1-2/\eta}}{\eta}\Gamma\left(\frac{2}{\eta}\right)
\zeta\left(\frac{2}{\eta}\right)
\end{equation}
where $Q_{gl}$ is the number of gapless points in the MBZ.
Here $\zeta(x)$ is Riemann's zeta function and $\Gamma(x)$ is the gamma function.
 For a linear dispersion around the gapless points we especially obtain
\begin{equation} 
\alpha({\mathbf{q}},\omega)_{gl}=Q_{gl}\frac{m\omega^2\pi(k_BT)^2}{12\hbar^2\omega_c q
\gamma^2}.
\end{equation}
  The relative size of the contributions from the gapped and gapless parts of the spectrum 
is determined 
by $\langle\Delta({\mathbf{k}})\rangle/k_BT$,  ${\mathcal{F}}$ and $\gamma$. For 
 $\langle\Delta \rangle \,\raisebox{-0.4ex}{$\stackrel {>}{\sim}$}\,  3k_BT$ the 
contribution from the gapped part can be ignored and the attenuation is given by 
Eq.(\ref{gapless}). Since the normal state attenuation is given by 
 $\alpha({\mathbf{q}},\omega)_{0,N}=m^2\omega^2/\hbar^2qh$ we get
\begin{equation} \label{rellw}
\frac{\alpha({\mathbf{q}},\omega)_{0,gl}}{\alpha({\mathbf{q}},\omega)_{0,N}}=
Q_{gl}\frac{2\hbar}{m\omega_c}\left(\frac{k_BT}{\gamma}
\right)^{2/\eta}\frac{1-2^{1-2/\eta}}{\eta}\Gamma\left(\frac{2}{\eta}\right)
\zeta\left(\frac{2}{\eta}\right).
\end{equation}
This calculation is valid in the clean case where the momentum is conserved during the 
scattering process. As a first approximation to the dirty limit we can assume that there 
is no momentum conservation in the scattering process and take $k_z$ and $k_z'$ as 
free variables~\cite{Mahan} (i.e $k_z'\neq k_z+q$). As in the Meissner state, this relaxation 
does not alter the result stated in Eq.(\ref{rellw}). This is due to the fact we are looking at 
energies close to the Fermi level such that we can assume that the normal state 
density-of-states is constant. It has been shown~\cite{Dukan2} that for small impurity
 concentrations and 
weak scattering potentials the density of states behaves essentially in the same way as for
 the pure case. We therefore believe that our results are somewhat insensitive to the presence 
of impurities. 

The Landau level structure of the normal state electron energies implies  that there will 
oscillations in the attenuation as the external field varies. The oscillatory terms are given 
by the  $j\neq 0$ terms in Eq.(\ref{pois}).  We end up with the following expression for the first
 harmonic  $\alpha({\mathbf{q}},\omega)_{gl,1}$ of the attenuation from the gapless part of the
 spectrum:
\begin{eqnarray}
\alpha({\mathbf{q}},\omega)_{gl,1}=Q_{gl}\frac{2m\omega^2}{\hbar k_BTq\pi}
\cos(2\pi(n_f-\epsilon(k_z^*)/\hbar\omega_c))\int_0^{\infty}kdk\int_0^{\infty}dx\nonumber \\
\times
(e^{-\sqrt{x^2\hbar^2\omega_c^2+\gamma^2k^{2\eta}}/k_BT}+1)^{-1}
(e^{\sqrt{x^2\hbar^2\omega_c^2+\gamma^2k^{2\eta}}/k_BT}+1)^{-1}\frac{\cos(2\pi x)x\omega_c}
{\sqrt{x^2\hbar^2\omega_c^2+\gamma^2k^{2\eta}}}
\end{eqnarray}

This integral can be solved in the case of a linear spectrum around the gapless points 
(i.e $\eta=1$). In this case we obtain for the 1st harmonic of the attenuation
\begin{eqnarray} \label{gapless1}
\alpha({\mathbf{q}},\omega)_{gl,1}=Q_{gl}\frac{m\omega^2}{q\hbar^2\omega_c\pi}
\left(\frac{k_BT}{\gamma}\right)^2\cos(2\pi(n_f-\epsilon(k_z^*)/\hbar\omega_c)) \nonumber \\
\times \left[ \frac{\pi^2\cosh(2\pi ^2k_BT/\hbar\omega_c)}
{\sinh^2(2\pi^2k_BT/\hbar\omega_c)}-\frac{1}{(2\pi k_BT/\hbar\omega_c)^2}
\right].
\end{eqnarray}
For low temperatures this goes as $T^2$. In the case of a general dispersion law around the 
gapless point given by $\Delta({\mathbf{k}})= \gamma k^{\eta}$, the leading term for the 
1st harmonic goes as $(T/\gamma)^{2/\eta}$. This should be contrasted  to the Meissner 
state, or the contribution from the gapped part of the spectrum, where the oscillatory terms
  will again be damped by 
a factor $2/(\exp(\langle\Delta({\mathbf{k}})\rangle/k_BT)+1)$. In the case of the normal
 state, the 1st harmonic is
\begin{eqnarray}\label{normal1} \alpha({\mathbf{q}},\omega)_{N,1}=
\frac{m^2\omega^2}{hq\hbar^3\omega_c}\frac{4k_BT\pi^2}
{\sinh(2\pi^2k_BT/\hbar\omega_c)}\cos(2\pi(n_f-\epsilon(k_z^*)/\hbar\omega_c))
\end{eqnarray}
which is independent of $T$ for low temperatures.
In the case of a dispersion law along the $z$-direction given by 
 $\epsilon(k_z)=t\cos(k_za_z)$ we have to substitute $m/q$ by 
 $2/ta_z|\sin(k_z^*a_z)-\sin((k_z^*+q)a_z)|$ in Eq.(\ref{gapless}) and 
 $m\cos(2\pi(n_f-\epsilon(k_z^*)/\hbar\omega_c))/q$ by 
 $\sum_{i=1,2}\cos(2\pi(n_f-\epsilon(k_{z,i}^*)/\hbar\omega_c))/ta_z|\sin(k_z^*a_z)-\sin((k_z^*+q)a_z)| $
 in Eq.(\ref{gapless1})-(\ref{normal1}) where $k_{z,i}$ are the two solutions of the 
equation $t\cos((k_z+q)a_z)-t\cos(k_za_z)=\hbar \omega$.

In Fig. 2 we show an example of the acoustic attenuation calculated numerically  
for two different coupling strengths as a function of $n_f$. We have solved the BdG 
equations self-consistently in 3D as a function of  the external magnetic field at constant  
chemical potential. In this paper we have chosen to work with a constant chemical potential.
 The difference 
between holding the chemical potential constant and holding the number of particles constant 
is negligible in 3D for the normal state.~\cite{Schoenberg} Even in 2D it can be shown that the 
superconducting order tends to suppress the difference between the two 
cases in the mixed state.~\cite{Bruun} The attenuation is  calculated via Eq.(\ref{master}). 
We have chosen parameters such that 
 $\omega_D/\omega_c=5$,   $k_BT/\hbar\omega_c=0.05$ and $\omega/\omega_c=0.01$ 
when $n_f=12$, where $\omega_D$ is the usual cutoff of the pairing interaction around the Fermi 
surface. The solid curve is the normal state attenuation which is continued into the 
mixed state to facilitate comparison with the mixed state attenuation; the dashed curve
corresponds to the coupling strength $g/\hbar\omega_c l^3=7.85$
 while the dash-dot curve corresponds to $g/\hbar\omega_c l^3=8.7$. 
The dispersion law along the $z$-direction is $\epsilon(k_z)=t\cos(k_za_z)$ where 
 $t/\hbar\omega_c=0.4$. In Fig. 3 we have plotted the corresponding order-parameter 
which in the lowest Landau level approximation for the pairing 
(which is valid close to $H_{c2}$)~\cite{Rasolt} can be 
characterized by the dimensionless number 
 $\Delta_0\propto  \Delta({\mathbf{k}})$. The connection between $\Delta_0$ and 
 $\Delta({\mathbf{k}})$ or $\Delta({\mathbf{r}})$ is given by an obvious generalisation 
to 3D of the results by Norman \textit{et al}.~\cite{Big Mac}
The phase transition between the normal and the mixed state 
occurs for  $n_f\simeq 7.6$ with  $g/\hbar\omega_c l^3=8.7$ and for $n_f\simeq 9.5$ with 
 $g/\hbar\omega_c l^3=7.85$. As can be seen from Fig. 2, the oscillations of the attenuation
  due to the Landau level quantization persist into the mixed state, although 
they are damped as compared to the normal state oscillations. Eventually the oscillations die
 out when the off-diagonal pairing becomes dominant and the diagonal approximation 
breakes down. This happens for  $n_f\,\raisebox{-0.4ex}{$\stackrel {>}{\sim}$}\,11$ with
$g/\hbar\omega_c l^3=8.7$.
Comparing the attenuation for the two coupling strengths in the region  
 $10\le n_f \le 11$ we get $\alpha_0(7.85)/\alpha_0(8.7)\simeq 1.9$ and 
 $\alpha_1(7.85)/\alpha_1(8.7)\simeq 2.4$ where $\alpha(g)_i$ is the $i$'th harmonic 
 of the attenuation for the coupling strength $g$. Since  
 $\langle \Delta({\mathbf{k}}) \rangle$ can be calculated~\cite{Big Mac} directly from 
 $\Delta_0$, and the coefficient $\gamma$ in the dispersion law around the gapless points is
  proportional to $g\Delta_0$, we can compare the numerical 
results with the analytical predictions outlined above. If the quasiparticle spectrum is 
essentially gapped, Eq.(\ref{gapped}) predicts that  
 $\alpha_0(7.85)/\alpha_0(8.7)=f(\langle \Delta_{7.85}\rangle)
 /f(\langle \Delta_{8.7}\rangle)\simeq 8.8$.
 If the attenuation is primarily originating from gapless points in 
 the quasiparticle spectrum, Eq.(\ref{gapless}) predicts that 
 $\alpha_0(7.85)/\alpha_0(8.7)=(8.7\Delta_0(8.7)/7.85\Delta_0(7.85))^{2/\eta}$. This gives 
 $2.5$ for $\eta=1$ and $1.6$ for $\eta=2$. Hence the numerical calculation 
 $(\alpha_0(7.85)/\alpha_0(8.7)\simeq 1.9$) imply (i) that the 
gapless points dominate the attenuation in agreement with our analytical results and (ii) that 
the dispersion law is somewhere between $\eta=1$ and $\eta=2$, since the gapless
 predictions agree reasonably well with the numerical results while the gapped predictions are 
qualitatively wrong. We cannot 
however make a quantitative numerical determination of the dispersion law around the 
gapless points.  This is due to the fact that in 
order to reduce the computation load, which is high in this 3D case, we have to choose a 
 ${\mathbf{k}}$-mesh with a rather large spacing between the points. (the mesh consists of  
 $100\times 50$ points) This means that the gapless regions in the quasiparticle 
spectrum are only probed by a few ${\mathbf{k}}$-vectors, thereby prohibiting  a quantitative 
determination of the dispersion law. Likewise, the analysis above predicts that 
 $\alpha_1(7.85)/\alpha_1(8.7)$ equals $2.5$ and $1.6$ for $\eta=1$ and $\eta=2$ 
respectively.  The $\eta=1$ prediction of $2.5$ agrees well with the numerical result 
$\alpha_1(7.85)/\alpha_1(8.7)\simeq 2.4$. Again, any quantitative comparison would require 
a much finer ${\mathbf{k}}$-mesh. A further complication is that due to the  
number of  Landau levels participating in the pairing (approx. 10), the oscillations in the
 attenuation 
are quickly damped and the diagonal approximation is only valid over  relatively
 few oscillations. This problem could be avoided if we were to perform the calculations for 
experimentally realistic parameters, where there are many more Landau levels in the 
pairing region; however  we have not been able to run the programs for such parameters 
due to the intense computation load. The above example does show however that the
 numerical calculations support our analytic theory. In short, the normal state oscillations
 in the attenuation  continue into the mixed state and  the damping 
is dominated by gapless points not too far into the mixed state.

 Hence we have calculated the 0'th and the 1st harmonic of the acoustic attenuation in the
 mixed state. The presence of gapless points enhances the acoustic attenuation above the 
conventional value for the Meissner state. 
When $\langle\Delta \rangle \,\raisebox{-0.4ex}{$\stackrel {>}{\sim}$}\,  3k_BT$ such
 that we can ignore the contribution from the gapped part of the spectrum, the temperature 
dependence of the attenuation is a power law given by $\alpha \propto T^{2/\eta}$. 
Furthermore, we predict that one 
should observe oscillations in the signal as the external field is varied. The magnitude of 
these oscillations should have the same temperature dependence as the average value of the 
signal. Hence, by looking at the temperature dependence of the
 attenuation one should be able to detect the presence of the gapless points and to extract
 the dispersion law around these points.
From the diagonal approximation it follows that $\gamma \propto \langle \Delta \rangle$ 
where  $\langle \Delta \rangle$ is the ${\mathbf{k}}$-space (or real space) average of the 
gap. Hence close to the upper critical field $H_{c2}$ we expect 
 $\gamma \propto \sqrt{1-H/H_{c2}}$. This would mean that 
 $\alpha({\mathbf{q}},\omega)\propto (1-H/H_{c2})^{-1/\eta}$  close to $H_{c2}$.

\section{Low Temperature}
We will now consider the limit where $k_B T\ll \hbar \omega$. In this limit  
 only the third delta function in Eq.(\ref{master}), which describes the creation of 
two quasiparticles will contribute to the damping. We first calculate the
 0'th harmonic of the attenuation. 
 Using the Poisson identity, making a substitution of variables and using the fact that 
 $v_s\equiv \omega/q\ll v_f$ ($v_f$ is the Fermi velocity) 
we obtain  the 0'th harmonic of the attenuation, when $\epsilon(k_z)=k_z^2/2m$, as
\begin{equation}
\alpha({\mathbf{q}},\omega)_0=\frac{m\omega}{q\hbar^3\omega_c L_xL_y}\sum_
{{\mathbf{k}}}\int_{\Delta({\mathbf{k}})}^{\omega-\Delta({\mathbf{k}})}
dE[1-f(E)-f(\omega-E)]
\frac{E(\omega-E)+\Delta^2}{\sqrt{(E^2-\Delta^2)((\omega-E)^2-\Delta^2)}}.
\end{equation} 
In the limit $\omega/ k_B T \rightarrow \infty$ this integral can be written as a 
complete elliptic integral~\cite{Bobetic} and we obtain
\begin{equation} \alpha({\mathbf{q}},\omega)_0=\frac{m\omega^2}{4\pi^2\hbar^2
\omega_c q}
\int d^2{\mathbf{k}}{\mathcal{E}}\left(\sqrt{1-4\Delta({\mathbf{k}})^2/\hbar^2\omega^2}\right).
\end{equation}
Here ${\mathcal{E}}(k)$ is the complete elliptic integral of the second kind.
 The existence of the gapless points again gives rise to a qualitatively different result for the
 ultrasonic 
attenuation in the mixed state as compared to the Meissner state. This is most easily 
understood  by the observation that there will always be attenuation for any frequency in
 the mixed state since 
there are always quasiparticle states with $\Delta({\mathbf{k}})\le \hbar\omega/2$. Hence the 
phonon will always have enough energy to create two quasiparticles. This is in 
contrast to the Meissner state, where there is no attenuation for 
 $\hbar\omega < 2 \Delta$.~\cite{Bobetic} Thus a direct experimental signature of these
 gapless points would be the absence of the discontinuity in the attenuation which is present 
in the Meissner state~\cite{Bobetic}, when 
 $\hbar\omega=2\Delta$, and the presence of acoustic 
attenuation in the mixed state as $\omega \rightarrow 0$. We again assume that the 
dispersion law around the gapless points to leading order is given by 
$\Delta({\mathbf{k}})=\gamma k^\eta$ 
($k=|{\mathbf{k}}|$) in the region that contributes to the attenuation (i.e for 
 $\Delta({\mathbf{k}})\le\hbar\omega/2$). Using this we obtain
\begin{equation} \label{lowT}
\alpha({\mathbf{q}},\omega)_0=Q_{gl}\frac{m \omega^{2/\eta+2}}{2\pi\hbar^2\omega_c 
q(2\gamma)^{2/\eta}}\int_0 ^1 xdx{\mathcal{E}}(\sqrt{1-x^{2\eta}})=n
\frac{m \omega^{2/\eta+2}}{2\pi\hbar^2\omega_c q(2\gamma)^{2/\eta}}I_{\eta}.
\end{equation}
 Since the attenuation 
 for the normal state is $\alpha({\mathbf{q}},\omega)_{0,N}=m^2\omega^2/2\pi\hbar^3q$
 we have
\begin{equation} \label{relT0}
\frac{\alpha({\mathbf{q}},\omega)_0}{\alpha({\mathbf{q}},\omega)_{0,N}}=Q_{gl}
\left(\frac{\omega}{2\gamma}\right)^{2/\eta}\frac{\hbar I_{\eta}}{m\omega_c}.
\end{equation}
   For the same reasons as for the $\hbar \omega \ll k_BT$ case, we expect Eq.(\ref{relT0}) 
to also be valid in the dirty limit. The remaining integral in Eq. (\ref{lowT}) can be solved 
for various $\eta$. We obtain, for instance, $I_1=2/3$, $I_2=\pi^2/16$, $I_{1/2}=32/45$ etc. 
Again the attenuation has an oscillatory behaviour as a function of the 
magnetic field $H^{-1}$. From the Poisson identity we get for the 1st harmonic
\begin{eqnarray}
\alpha({\mathbf{q}},\omega)_1=Q_{gl}\frac{m\omega}{4\hbar^3\omega_cq2\pi}\sum_{j=-1,1}
e^{2\pi i j(n_f-\epsilon(k_z^*)/\hbar\omega_c)}\int kdk\int d\xi d\xi'\delta(E+E'-\omega)\nonumber\\
\times \left[ (1-\xi/E)(1+\xi'/E')+\frac{\Delta(k)^2}{EE'}\right]e^{ij\xi/\hbar\omega_c}.
\end{eqnarray}
We have not been able to solve this integral exactly. However, in the region where the 
diagonal approximation holds (i.e for  
 $\omega\,\raisebox{-0.4ex}{$\stackrel {<}{\sim}$}\,\hbar\omega_c/4$) one can  expand the
 factor $e^{il\xi/\hbar\omega_c}$ to a good approximation. This immediately yields that, to leading
 order in $\omega/\omega_c$, the amplitude of the first harmonic varies as 
 $\omega^{2/\eta+2}$.
The next correction term will go as $\omega^{2/\eta+4}$. 

So we see that the existence of the gapless points in the MBZ implies that there is a finite 
attenuation for any frequency of the sound wave. There will be no discontinuity in the 
attenuation as a function of the sound wave frequency. As the external field is varied 
one should observe oscillations in the attenuation. The dependence of the attenuation on 
frequency is algebraic and the
 power-law is determined by the dispersion law around the gapless points. 
If $\Delta({\mathbf{k}})=\gamma k^\eta$ we obtain  $\alpha \propto \omega^{2/\eta+2}$. 
The absence of the discontinuity and the frequency dependence of the attenuation should  
in principle provide the possibility of experimentally determining the existence and dispersion 
law for the gapless points. Again we expect 
 $\alpha({\mathbf{q}},\omega)\propto (1-H/H_{c2})^{-1/\eta}$  close to $H_{c2}$. 
By making the same substitutions as in the $\hbar \omega \ll k_BT$ limit, one can obtain the 
results for the case $\epsilon(k_z)=t\cos(k_za_z)$ relevant for layered structures.

\section{Conclusion}
In this paper we have considered the acoustic attenuation in a type II BCS superconductor at
  high magnetic fields using both numerical and analytical methods. We have shown that 
away from the semiclassical regime 
where the Landau level structure of the electronic states is important, the attenuation 
will in general be an oscillatory function of the external magnetic field. Furthermore, since
 the attenuation probes the quasiparticle density of states, the  presence of 
gapless points in the 
quasiparticle spectrum makes the attenuation qualitatively different as compared to the 
Meissner state attenuation. For  $k_BT\ll \hbar \omega_c$ the attenuation is an algebraic function of the
 frequency and there is no discontinuity as opposed to the Meissner state attenuation. For 
$k_BT\gg \hbar \omega_c$ the attenuation is an algebraic function of the
 temperature. The exponent of the power law is determined by the dispersion law around the 
gapless points. This behaviour should in principle be experimentally detectable; such an 
experiment would provide confirmation of the existence and nature of the gapless points.
\section{Acknowledgements}
 This work has been supported in part by EPSRC grant
GR/K 15619 (VNN and NFJ) and by The Danish Research Academy (GMB).

\newpage
\begin{center}{Figure Captions}\end{center}
\bigskip
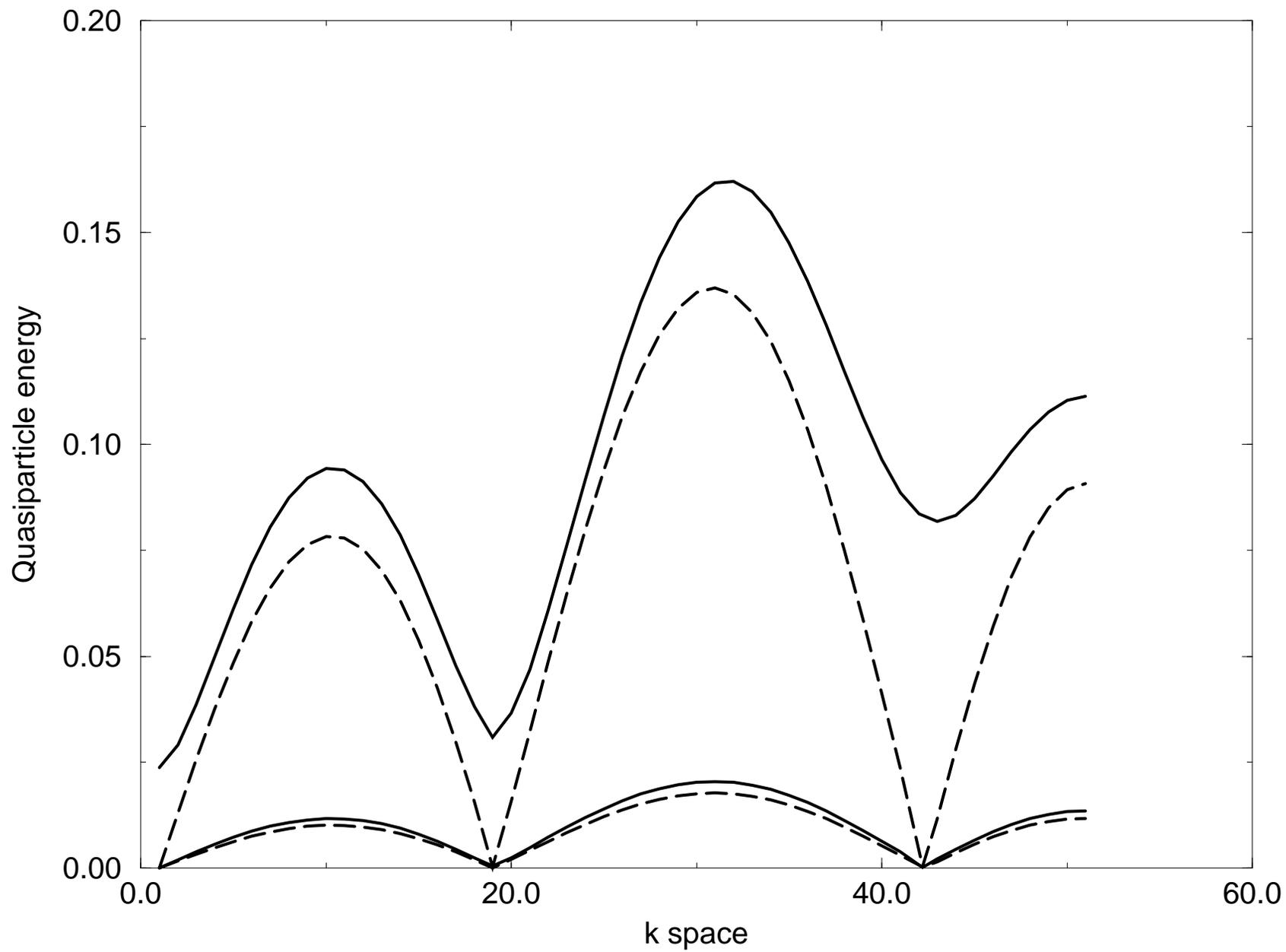
\noindent Fig.\ 1: The lowest quasiparticle energy in units of $\hbar \omega_c$ for two 
different values of the order parameter 
 ($\langle \Delta ({\mathbf{k}}) \rangle\simeq 0.3\hbar \omega_c$ and 
  $\langle \Delta ({\mathbf{k}}) \rangle\simeq 0.05\hbar \omega_c$). The 
 ${\mathbf{k}}$-vector is measured in units of $2\pi/L_x$ where $L_x$ is the size of the sample 
in the $x$-direction. The solid lines show the exact numerical result while the dashed lines 
correspond to the diagonal approximation.

\

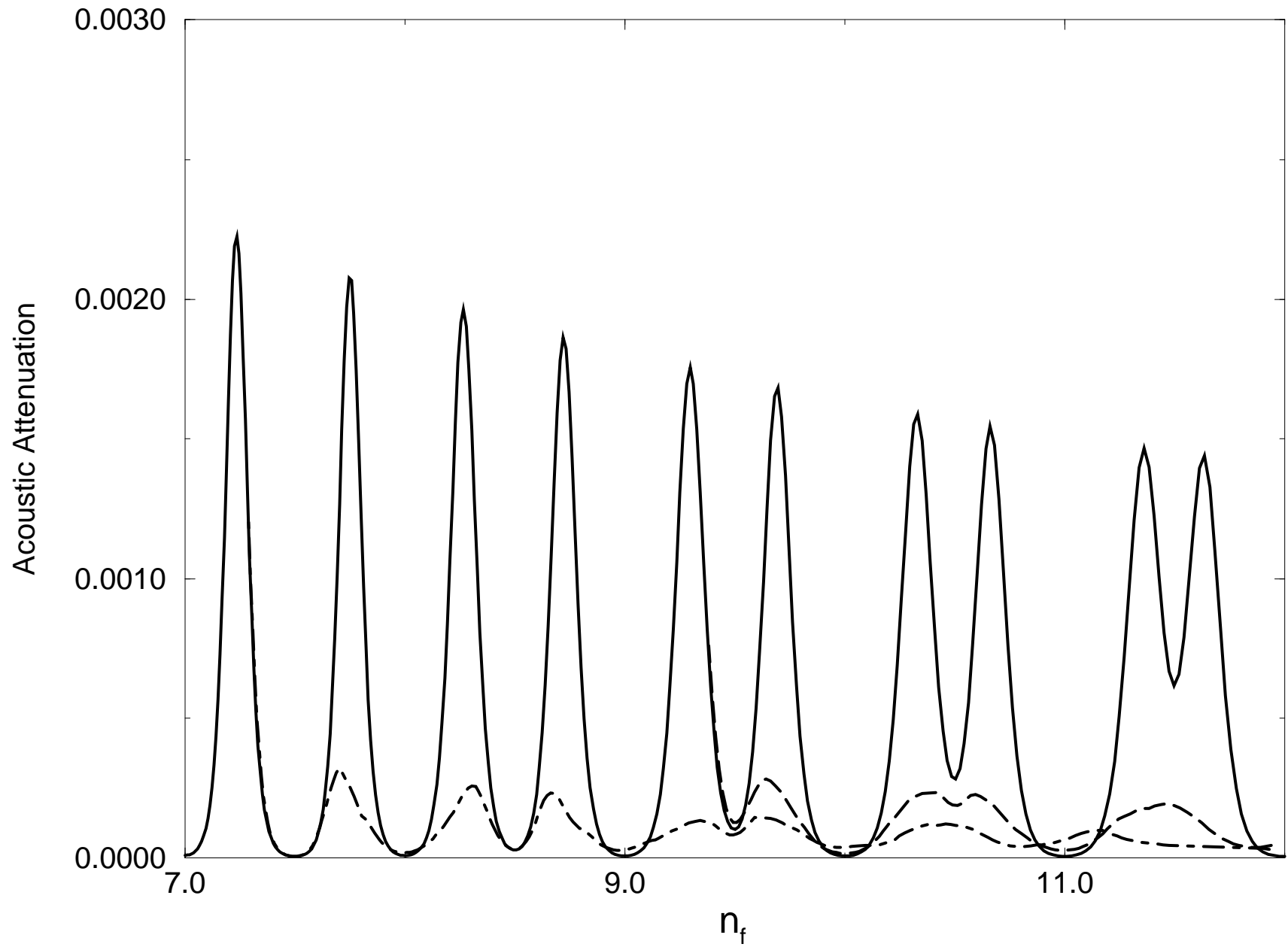
\noindent Fig.\ 2: The attenuation as a function of $n_f$. The solid line is the normal state 
attenuation. The dashed line is for the coupling $g/\hbar\omega_c l^3=7.85$.  The 
dash-dot line is for $g/\hbar\omega_c l^3=8.7$.

\
 
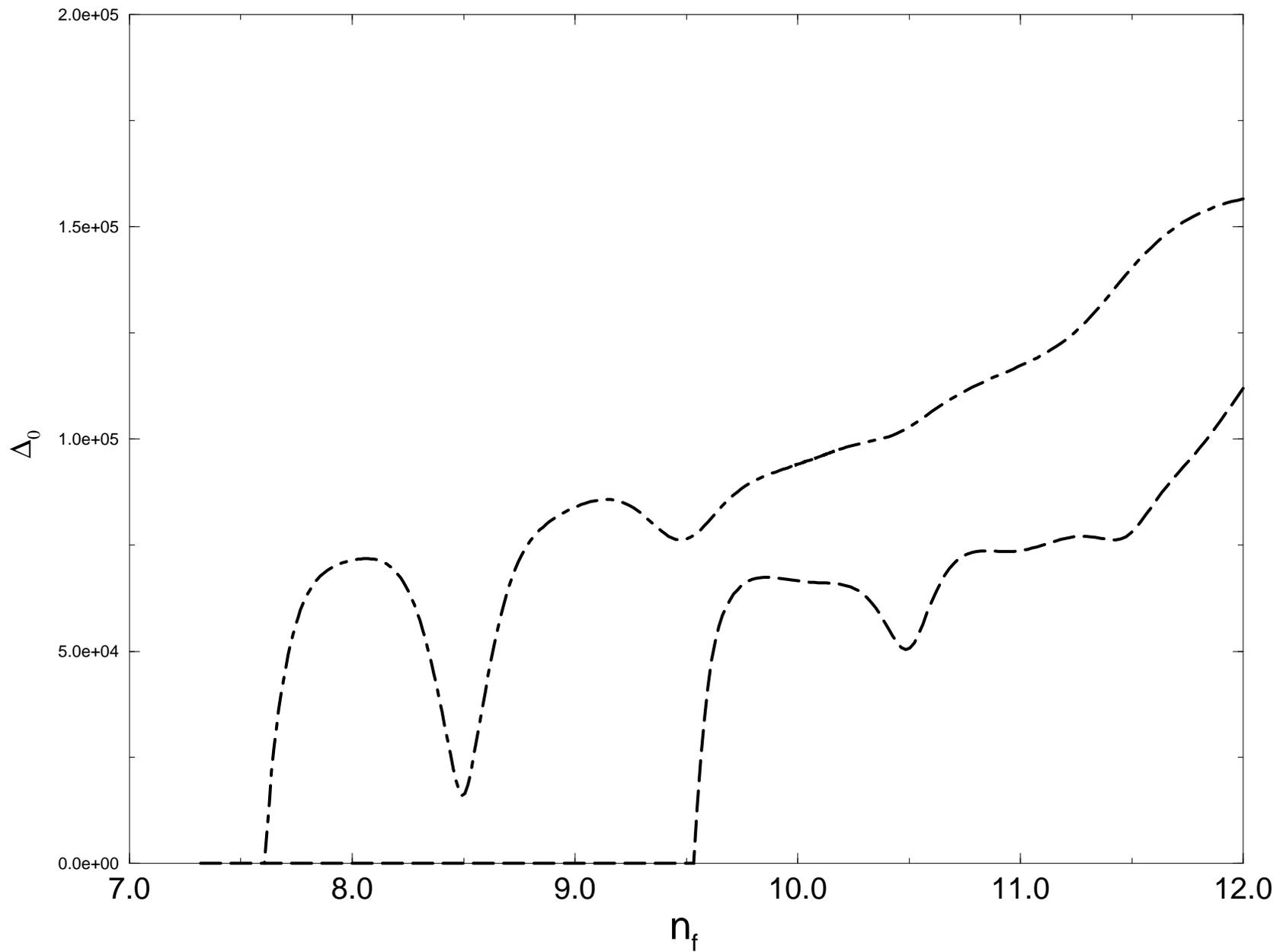
\noindent Fig.\ 3: The order-parameter as a function of $n_f$. The dashed line is for the 
coupling $g/\hbar\omega_c l^3=7.85$ while the dash-dot line is 
for $g/\hbar\omega_c l^3=8.7$.
\end{document}